\documentclass[sigconf,screen,]{acmart}

\usepackage{cite}
\usepackage[ruled, linesnumbered]{algorithm2e}

\usepackage{amsmath,amssymb,amsfonts}
\usepackage{url}
\usepackage{booktabs}
\usepackage{xurl}
\usepackage{caption}
\usepackage{subcaption}
\usepackage{multirow}
\usepackage{multicol}
\usepackage{algorithmic}
\usepackage{graphicx}
\usepackage{textcomp}
\usepackage{xcolor}
\usepackage{color}
\usepackage{listings}
\usepackage{xspace}
\usepackage{booktabs}
\usepackage{array}
\usepackage{enumitem}
\usepackage{tcolorbox}

\usepackage{bbm}

\newcommand{\finding}[2]{
\begin{tcolorbox}[width=\linewidth,boxrule=0pt,top=1pt, bottom=1pt, left=1pt,right=1pt, colback=gray!20,colframe=gray!20]
\textbf{Finding #1:} 
{#2}
\end{tcolorbox}}

\copyrightyear{2025} 
\acmYear{2025} 
\setcopyright{cc}
\setcctype{by}
\acmConference[FSE Companion '25]{33rd ACM International Conference on the Foundations of Software Engineering}{June 23--28, 2025}{Trondheim, Norway}
\acmBooktitle{33rd ACM International Conference on the Foundations of Software Engineering (FSE Companion '25), June 23--28, 2025, Trondheim, Norway}
\acmDOI{10.1145/3696630.3728535}
\acmISBN{979-8-4007-1276-0/2025/06}

\begin{document}

\title{RAG or Fine-tuning? A Comparative Study on LCMs-based Code Completion in Industry}

\author{Chaozheng Wang}
\orcid{0000-0002-3935-7328}
\affiliation{%
  \institution{Chinese University of Hong Kong}
  \city{Hong Kong}
  \country{China}
}
\email{adf111178@gmail.com}

\author{Zezhou Yang}
\orcid{0009-0008-9092-3381}
\affiliation{%
  \institution{Tencent}
  \city{Guangzhou}
  \country{China}
}
\email{zezhouyang@tencent.com}

\author{Shuzheng Gao}
\orcid{0000-0002-8102-480X}
\affiliation{%
  \institution{Chinese University of Hong Kong}
  \city{Hong Kong}
  \country{China}
}
\email{szgao23@cse.cuhk.edu.hk}

\author{Cuiyun Gao}
\orcid{0000-0001-8513-6836}
\authornote{Cuiyun Gao is the corresponding author.}
\affiliation{%
  \institution{Chinese University of Hong Kong}
  \city{Hong Kong}
  \country{China}
}
\email{cuiyungao@outlook.com}

\author{Ting Peng}
\orcid{0009-0003-6970-0857}
\affiliation{%
  \institution{Tencent}
  \city{Guangzhou}
  \country{China}
}
\email{sakurapeng@tencent.com}

\author{Hailiang Huang}
\orcid{0009-0004-0655-9398}
\affiliation{%
  \institution{Tencent}
  \city{Guangzhou}
  \country{China}
}
\email{eraserhuang@tencent.com}

\author{Yuetang Deng}
\orcid{0009-0003-7060-4109}
\affiliation{%
  \institution{Tencent}
  \city{Guangzhou}
  \country{China}
}
\email{yuetangdeng@tencent.com}

\author{Michael Lyu}
\orcid{0000-0002-3666-5798}
\affiliation{%
  \institution{Chinese University of Hong Kong}
  \city{Hong Kong}
  \country{China}
}
\email{lyu@cse.cuhk.edu.hk}




\renewcommand{\shortauthors}{Wang et al.}

\begin{abstract}
Code completion, a crucial practice in industrial settings, helps developers improve programming efficiency by automatically suggesting code snippets during development. With the emergence of Large Code Models (LCMs), this field has witnessed significant advancements. Due to the natural differences between open-source and industrial codebases, such as coding patterns and unique internal dependencies, it is a common practice for developers to conduct domain adaptation when adopting LCMs in industry. There exist multiple adaptation approaches, among which retrieval-augmented generation (RAG) and fine-tuning are the two most popular paradigms. However, no prior research has explored the trade-off of the two approaches in industrial scenarios.


To mitigate the gap, we comprehensively compare the two paradigms including Retrieval-Augmented Generation (RAG) and Fine-tuning (FT), for industrial code completion in this paper. In collaboration with Tencent's WXG department, we collect over 160,000 internal C++ files as our codebase. We then compare the two types of adaptation approaches from three dimensions that are concerned by industrial practitioners, including effectiveness, efficiency, and parameter sensitivity, using six LCMs. Our findings reveal that RAG, when implemented with appropriate embedding models that map code snippets into dense vector representations, can achieve higher accuracy than fine-tuning alone. Specifically, BM25 presents superior retrieval effectiveness and efficiency among studied RAG methods. Moreover, RAG and fine-tuning are orthogonal and their combination leads to further improvement. We also observe that RAG demonstrates better scalability than FT, showing more sustained performance gains with larger scales of codebase. 
Our findings provide actionable guidance for choosing and implementing appropriate methods to adopt LCMs based on specific industrial scenarios and requirements.

\end{abstract}



\maketitle

\section{Introduction}
Code completion stands as a fundamental and critical task in code intelligence, aiming to predict subsequent code tokens or statements that developers intend to write. By anticipating developers' coding intentions, code completion remarkably enhances programming productivity and has become an indispensable feature in modern Integrated Development Environments (IDEs)~\citep{copilot}. The advent of Large Language Models (LLMs)~\citep{DBLP:conf/nips/BrownMRSKDNSSAA20} has led to the emergence of Large Code Models (LCMs)~\citep{hui2024qwen2,guo2024deepseek}, which bring substantial improvements to code completion performance~\citep{hui2024qwen2}. This advancement has catalyzed the development of multiple commercial tools~\citep{cursor,copilot}, offering personalized code completion services. 

Despite LCMs' impressive performance in general development scenarios such as utility scripts and application development, their effectiveness substantially deteriorates when applied to industrial proprietary codebases \citep{ahmed2024studying}. This performance gap arises from the inherent complexity and massive scale of industrial code, which often undergoes extensive business-specific implementation. For instance, our industrial partner Tencent WXG operates WeChat, China's largest instant messaging platform with over one billion monthly active users. Beyond basic communication services, the platform supports a rich ecosystem of features including social networking (Moments) and short-form videos. To accommodate this massive user base and diverse functionality demands, WXG has developed an extensive and sophisticated backend infrastructure with unique architectural patterns and optimizations. These specialized implementations significantly deviate from conventional open-source solutions, creating a substantial distributional shift between industrial code patterns and the open-source training data used by LCMs. Moreover, the sensitive nature of proprietary code and associated privacy concerns preclude the direct utilization of commercial code completion solutions, even those with superior capabilities, necessitating alternative approaches for industrial code completion.

Given the distribution gap between industrial and open-source codebase, two predominant approaches have emerged to adapt LCMs to new domains: Retrieval-Augmented Generation (RAG)~\citep{DBLP:conf/acl/LuDHGHS22,DBLP:conf/sigsoft/Wang0JH23} and fine-tuning (FT)~\citep{DBLP:conf/sigsoft/WangYGP0L22,DBLP:conf/kbse/LiZCHD24}. Specifically, given a novel source code repository previously unseen by the model, RAG enhances generation by retrieving relative code snippets from the repository as prior knowledge. This retrieval-based augmentation manifests in two forms: (1) prompt-based augmentation \citep{xu2024chatqa}, where retrieved similar code segments are prepended to the input query, providing contextual examples; and (2) logits-based augmentation \citep{tang2023domain, khandelwal2019generalization}, where the distribution of similar usage patterns and outputs serves as a prior to calibrate the model's prediction distribution. In contrast, FT directly optimizes the model's parameters on the target codebase, aiming at aligning the model's internal distribution with the domain-specific patterns present in
industrial repositories.


While both RAG and FT approaches have demonstrated effectiveness across diverse domains, such as document QA and
code intelligence~\citep{DBLP:conf/acl/LuDHGHS22,DBLP:conf/kbse/LiZCHD24,DBLP:journals/tse/WangYGPZL23,DBLP:conf/sigsoft/Wang0JH23}, these studies predominantly focus on public datasets containing general knowledge. Industrial codebases present unique challenges with their complex dependencies, company-specific patterns, and proprietary frameworks, making the direct application of existing findings uncertain. The efficacy of these approaches in such specialized, real-world environments remains largely unexplored. 
Existing RAG-based code completion research~\citep{DBLP:journals/corr/abs-2407-19487,DBLP:conf/iclr/0003XM24} has primarily focused on Python and Java codebases, where implementations tend to be more concise and less intricate than C++ systems which are widely used in Tencent WXG \citep{wilkie1998measuring, prechelt2000empirical}. Moreover, these studies have been conducted on relatively small-scale open-source repositories, making their findings potentially less applicable to industrial environments. The massive scale of industrial codebases significantly increases retrieval complexity, as the search space expands exponentially and the identification of relevant code snippets becomes more challenging. For instance, logits-based retrieval methods~\citep{DBLP:conf/emnlp/ZhangCZKLZMLC23, tang2023domain} provide feasible and effective solutions for small codebases. However, they become prohibitively expensive when scaling to industrial repositories with hundreds of thousands of files, as they require storing and searching through massive logit matrices for each code segment. These distinctive characteristics of industrial codebases necessitate a thorough investigation of existing adaptation approaches tailored to these specific challenges.

In this paper, we conduct a comprehensive empirical study comparing RAG and FT
approaches for adapting LLMs to industrial code completion tasks. We use the backend codebase with more than 160,000 files of Tencent WXG as our experimental testbed, which contains various business functionalities and represents the complexity typical of industrial software systems. 
To systematically investigate adaptation approaches, we construct a subset of 120,000 C++ files as both the retrieval corpus for RAG and the training data for FT experiments. For the RAG approaches, we explore two main categories:
(1) similarity-based retrieval~\citep{DBLP:conf/emnlp/ZhangCZKLZMLC23}, which identifies syntactically or semantically similar code snippets, and (2) dependency-based retrieval~\citep{DBLP:conf/kbse/LiuYZS00JW24}, which leverages project-specific dependency relationships and usage patterns. Our evaluation comprehensively compares these RAG variants against FT approaches across multiple dimensions, including effectiveness, efficiency, and parameter sensitivity.


Through extensive experiments, our study reveals several key findings in enhancing LCM-based industrial code completion. RAG and FT substantially improve LCMs' performance, with RAG achieving a higher performance ceiling. Notably, among the evaluated RAG approaches, BM25 demonstrates superior effectiveness while maintaining computational efficiency comparing with neural embedding models. The combination of RAG and FT yields further performance gains, suggesting their complementary nature. Regarding computational costs, FT primarily requires resources during the preparation stage for model training, while RAG's overhead mainly occurs during inference as it needs to retrieve and process relevant code snippets in runtime. Furthermore, we observe that RAG demonstrates better scalability than FT, showing more sustained performance gains with larger scales of codebase.

Our key contributions are summarized as follows:
\begin{itemize}
\item We present the first comprehensive study of LCM adaptation approaches on a large-scale proprietary industrial codebase. Our evaluation reveals the performance limitations of current state-of-the-art LCMs when directly applied to industrial code completion tasks, highlighting the necessity for more effective domain adaptation.

\item We conduct extensive experiments comparing RAG and FT approaches across multiple dimensions, including effectiveness, efficiency, and data scaling. This systematic evaluation provides a detailed understanding of the trade-offs between different adaptation approaches in industrial settings.

\item We derive practical insights and guidelines for developers seeking to adapt LCMs for code completion in proprietary codebases, offering empirically validated recommendations for implementing appropriate adaptation approaches.

\end{itemize}
\section{Research Methodology}

\subsection{Data Preparation}
\begin{table*}[th]
    \centering
    \caption{Statistics of our dataset, where the number of tokens are calculated by the tokenizer of DeepSee-Coder.}
    \vspace{-4pt}
    \begin{tabular}{c|cccccc}
    \toprule
      Split & \# Files & \# Tokens & \# LOC& \# Instances & Avg. Input Token & Avg. Target Token \\
    \midrule
      Train   & 120,000 & 401M & 25.5M&-  & - & -\\
      Validation & 23,746 & 69M & 3.8M & -  & - & -\\
     Test &  20,000 & 63M & 3.5M& 50,000 & 246.75 & 17.01 \\
     \bottomrule
    \end{tabular}
    \vspace{-6pt}
    \label{tab:statis}
\end{table*}

\subsubsection{Data Collection and Filtering}
We collect a comprehensive snapshot of our backend C++ codebase from our department's repository. To ensure data quality and minimize potential training artifacts, we apply the following filtering criteria:

\begin{itemize}
\item Remove duplicate files and auto-generated framework code (i.e., compilation files) to prevent over-fitting of templated patterns
\item Filter out files containing extensive Chinese comments to maintain consistency in the training corpus
\item Eliminate files with unusually long C++ definitions (\# define), as these rare patterns could potentially lead to catastrophic repetition in model outputs.
\end{itemize}

After applying these filtering rules, we obtained a final dataset of 163,746 valid C++ source files. We further processed these files using ClangD~\citep{needle} to standardize code formatting, including consistent indentation and line break rules.

\subsubsection{Dataset Splitting}
To create a robust evaluation benchmark, we select 20,000 C++ files from our filtered dataset for generating input-output pairs. We employ a sliding window approach, where each input consists of 20 consecutive lines of code, and the target is the subsequent line. To ensure meaningful evaluation metrics, we apply additional filtering criteria to the target lines: (1) exclude instances where the target line contains only a single symbol, and (2) remove cases where the target line consists solely of comments.

After this process, we randomly sample 50,000 high-quality test instances, which form our evaluation benchmark for assessing model performance. The resulting test set covers various code completion scenarios while maintaining practical relevance to real-world development tasks.

For the remaining 143,746 source code files, we randomly select 120,000 for building retrieval database and fine-tuning LCMs, and 23,746 for validation. The statistics are shown in Table \ref{tab:statis}.

\subsection{Retrieval Augmented Generation}
\subsubsection{Retrieval Database Construction}

Direct utilization of our 120,000 C++ files for retrieval presents significant challenges. First, using entire C++ files as retrieval units is impractical due to their substantial length, with our codebase averaging over 3,300 tokens per file. Second, files often contain extraneous information that could potentially interfere with the model's inference process. 

To address these limitations, we decompose the files into fine-grained retrieval units. Specifically, we employ the Tree-Sitter~\citep{tree-sitter} to decompose each file into more granular units, specifically extracting individual functions and classes. We treat C++ files that either lack function definitions or contain only function declarations as single retrieval units. This decomposition strategy results in a final retrieval database of 914,667 items, providing a more focused and manageable collection of code segments. Specifically, we explore two kinds of RAG methods for code completion including similarity-based and dependency-based retrieval.

\subsubsection{Similarity-Based Retrieval}
Given our retrieval database $D$ containing 914,667 items, we implement a similarity-based retrieval mechanism. Let $\mathcal{M}$ denote the embedding model. For each item $d_i \in D$, we compute its embedding vector $\mathbf{v}_i = \mathcal{M}(d_i)$. These embeddings form our vector database $V = {\mathbf{v}_1, \mathbf{v}_2, ..., \mathbf{v}_n}$. During inference, for a given query $q$, we compute its embedding $\mathbf{q} = \mathcal{M}(q)$.



The similarity score between the query and each item in the database is calculated using cosine similarity: $\text{sim}(q, d_i) = \frac{\mathbf{q} \cdot \mathbf{v}_i}{|\mathbf{q}| |\mathbf{v}_i|}$.



We select the top-$K$ items with highest similarity scores: $d_{(1)}$, $d_{(2)}$, $...$, $d_{(K)}$, where $(i)$ denotes the index of the $i$-th most similar item. Following the work~\citep{gao2023makes}, these $K$ items are concatenated in ascending order of similarity and prepended to the query as examples. The augmented prompt is then fed to the model for generation:

\begin{equation}
\text{prompt} = d_{(K)} \oplus d_{(K-1)} \oplus ... \oplus d_{(1)} \oplus q
\end{equation}

\noindent where $\oplus$ denotes concatenation.

\subsubsection{Dependency-Based Retrieval}
Besides similarity-based retrieval, we also explore
a dependency-based retrieval method that leverages function call relationships to provide relevant context for code completion. Given a code completion query $q$, we first extract a sequence of function calls $C = [c_1, c_2, ..., c_m]$ using Tree-sitter~\citep{tree-sitter}, a robust parsing framework. For each function call $c_i$, we locate its corresponding function definition $d_i$ in the retrieval codebase, resulting in a sequence of function definitions $D = [d_1, d_2, ..., d_m]$.

The retrieved function definitions are then concatenated and prepended to the original query $q$ to form the augmented prompt:

\begin{equation}
\text{prompt} = d_m \oplus d_{m-1} \oplus ... \oplus d_1 \oplus q
\end{equation}

This approach ensures that the model has access to the implementation details of the functions being called in the query, providing context for more accurate code completion. By explicitly including function definitions, the model can better understand the expected behavior and usage patterns of the called functions.

\subsection{Fine-Tuning}
For the fine-tuning approach, we conduct training at the file level using source code. Following previous work~\citep{DBLP:conf/sigsoft/WangYGP0L22,DBLP:journals/tosem/GaoGHZNXL23,DBLP:conf/issta/WenGGXL24}, we first tokenize and concatenate the code text into a continuous sequence of tokens ${t_1, t_2, ..., t_n}$. Given a context length $L$, we segment the token sequence into non-overlapping blocks:

\begin{equation}
{(t_1, ..., t_L), (t_{L+1}, ..., t_{2L}), ..., (t_{n-L+1}, ..., t_n)}
\end{equation}

The model is then trained in an autoregressive manner to predict the next token given all previous tokens in the sequence:

\begin{equation}
\mathcal{L} = -\sum_{j=1}^{L-1} \log P(t_{j+1}|t_1, ..., t_j)
\end{equation}
\noindent where $\mathcal{L}$ denotes the loss function and $j$ means the token index.
\section{Experimental Setup}

\subsection{Selected LCMs}
In this paper, we select two kinds of popular and state-of-the-art LCMs with their versions in different sizes. In specific, our selected LCMs are:

\begin{itemize}
    \item \textbf{DeepSeek-Coder} (DSC) \citep{guo2024deepseek} is trained from 2T tokens from scratch, 
    achieving state-of-the-art performance in a variety of code intelligence tasks. Specifically, we choose DeepSeek-Coder Base in sizes of 1.3B and 6.7B in this paper.
    
    \item \textbf{Qwen2.5-Coder} (QC) \citep{hui2024qwen2} is a family of LCMs based on models of the Qwen2.5 series \citep{yang2024qwen2}. Qwen2.5 undergoes four-stage training with a total number of 5.5T tokens, presenting state-of-the-art performance in both code generation and reasoning. We choose 0.5B, 1.5B, 3B, and 7B versions of Qwen2.5-Coder.

\end{itemize}

\subsection{Evaluation Metrics}
Following previous work~\citep{DBLP:conf/kbse/WangGG0CGL24,lu2021codexglue,nashid2024contextual}, we select accuracy, edit similarity, and BLEU score to evaluate the effectiveness of different methods.

\subsubsection{Exact Match}
Exact match (EM) measures the percentage of predictions that exactly match the ground truth. For code completion, this is a strict metric that requires the model to generate exactly the same code line as the target.


\subsubsection{Edit Similarity}
Edit similarity (ES) measures how close the predicted code is to the ground truth based on the Levenshtein distance. It provides a more nuanced evaluation than accuracy by considering partial matches.


\subsubsection{BLEU Score}
BLEU (Bilingual Evaluation Understudy) score~\citep{DBLP:conf/acl/PapineniRWZ02} evaluates the quality of the generated code by comparing the n-gram overlap between the prediction and the ground truth.



\subsection{Embedding Models}
Following previous work \citep{gao2023makes}, we utilize the following retrieval methods for similarity-based RAG:
\begin{enumerate}
    \item \textbf{Random}: For each code completion query, we randomly select $K$ items from the retrieval database.

   \item \textbf{BM25} \citep{robertson2009probabilistic}: A classical information retrieval algorithm that ranks documents based on term frequency and inverse document frequency, treating code snippets as text documents, and computing relevance scores using keywords. BM25 has been widely used for code intelligence tasks~\citep{DBLP:conf/icse/ZhangW00020,DBLP:conf/icse/GaoMG000L24,gao2024search}.

    \item \textbf{CodeBERT} \citep{feng2020codebert}: A BERT-based pre-trained model specifically designed for programming language understanding, which we use to compute semantic embeddings of code snippets for similarity-based retrieval.
    
    \item \textbf{UniXcoder} \citep{guo2022unixcoder}: A unified cross-modal pre-trained model for programming language, which supports both natural language and programming language understanding tasks, serving as our embedding generator for code retrieval.
    
    \item \textbf{CodeT5} \citep{wang2021codet5}: An encoder-decoder model pre-trained on programming languages, which we utilize its encoder to generate embeddings for measuring code similarity in our retrieval system.

    \item \textbf{CoCoSoDa} \citep{shi2023cocosoda}: A contrastive learning-based code search approach that employs a soft data augmentation and momentum mechanism. It dynamically masks or replaces tokens with their types for input sequences and uses multimodal contrastive learning to align code-query pairs. 
\end{enumerate}
\subsection{Research Questions}
In this paper, we aim to answer four research questions.

\textbf{RQ1: How effective are RAG and fine-tuning approaches in industrial code completion?}
We evaluate the performance of directly applying LCMs to our code completion scenario and assess the improvements achieved through both RAG and fine-tuning approaches independently.

\textbf{RQ2: Can fine-tuning be effectively combined with RAG?}
We investigate whether combining fine-tuning with RAG can yield better performance than either approach alone, and analyze how fine-tuned models interact with the retrieval mechanism compared to base models.

\textbf{RQ3: How efficient are RAG and fine-tuning?}
We analyze the computational efficiency of each approach, measuring training and inference times, memory usage, and other resource requirements to understand their practical implications in industrial settings.

\textbf{RQ4: How do key parameters affect the performance of RAG and fine-tuning approaches?}
We conduct parameter sensitivity analysis from two aspects: (1) the impact of retrieved context size (Top-k) in RAG methods on model performance, and (2) the relationship between codebase scale and model effectiveness. 

\begin{table}[t]
    \centering
    \caption{Hyper-parameter settings.}
    \vspace{-6pt}
    \resizebox{\linewidth}{!}{
    {\begin{tabular}{c|c||c|c}
    \toprule
       Hyperparameter  & Value &  Hyperparameter  & Value\\
     \midrule
      Optimizer & AdamW\citep{loshchilov2018decoupled} & Warm-up steps   & 100 \\
      Learning rate & 5e-6 & Training batch size & 32 \\
      LR scheduler& Cosine Scheduler \citep{loshchilov2016sgdr}& Validation batch size & 32 \\
        Sequence Len. & 4,096 &  Adam epsilon & 1e-8 \\
      Max. gradient norm & 1.0 & Precision & BF16\\
      \midrule
      Max Gen. Tokens & 512 & Top-P & 0.95 \\
      \bottomrule
    \end{tabular}
    }
    }
    \vspace{-12pt}
    \label{tab:hyper}
\end{table}

\subsection{Implementation Details}
In the RAG framework, we implement two distinct retrieval strategies. For similarity-based retrieval, we leverage the BM25S library to construct and maintain a lexical-based retrieval database, while utilizing the Qdrant~\citep{qdrant} service to establish a vector-based retrieval system. For dependency-based retrieval, we first employ Tantivy~\citep{tantivy} to construct a mapping mechanism that links function names to their corresponding definitions. Subsequently, we utilize Tree-sitter \citep{tree-sitter} to analyze the code structure and extract function call dependencies, enabling us to retrieve the relevant function definitions based on the identified function names.

All the experiments are run on a server with Intel 8374C CPU and 8*A100 GPUs with 80GB of graphic memory. The hyper-parameter setting of the tuning procedure is listed in Table \ref{tab:hyper} following previous work. We enable the gradient checkpointing technique \citep{chen2016training} for LCMs to reduce the GPU memory consumption.

For fast inference, we utilize vLLM \citep{kwon2023efficient} based on PagedAttention to improve efficiency. To eliminate the influence of random sampling, we utilize greedy decoding strategy during inference. In addition, we employ the Flash-Attention technique~\citep{dao2022flashattention} for long-context optimization.
\section{Experiment Results}\label{sec:results}

\begin{table*}[t]
    \centering
    \caption{Results of different methods in the code completion task, where QC and DSC denote Qwen2.5-Coder and DeepSeek-Coder, respectively. The number of retrieved code snippets of RAG$_{sim}$ ($K$) is 5.}
    \vspace{-6pt}
    \begin{tabular}{l|cccccc|cc}
    \toprule
    Method & QC-0.5B & QC-1.5B& QC-3B & QC-7B & DSC-1.3B & DSC-6.7B&  Avg & Improve\\
    \midrule
    \multicolumn{9}{c}{Exact Match} \\
    \midrule
    Base & 21.09 & 24.48 & 25.89 & 28.04 & 23.16 & 26.07 & 24.79 & - \\
    Random & 20.40 & 23.72 & 25.09 & 27.31 & 22.52 & 25.89 & 24.16 & $\downarrow2.5\%$  \\
    RAG$_{Sim}$-BM25 & \textbf{52.52} & \textbf{55.81} & \textbf{57.13} & \textbf{58.08} & \textbf{48.90} & \textbf{50.13} & \textbf{53.76} & $\uparrow116.9\%$  \\
    RAG$_{Sim}$-CodeBERT & 20.48 & 23.79 & 25.29 & 27.20 & 22.36 & 25.86 & 24.16 & $\downarrow2.5\%$  \\
    RAG$_{Sim}$-UniXcoder & 38.39 & 41.47 & 43.16 & 46.99 & 38.98 & 40.46 &  41.64  &$\uparrow68.0\%$ \\
    RAG$_{Sim}$-CodeT5 & 39.05 & 42.38 & 43.91 & 45.42 & 40.37 & 42.93 & 42.34 &  $\uparrow70.8\%$ \\
    RAG$_{Sim}$-CoCoSoDa & 45.95 & 49.35 & 50.90 & 52.28 & 47.03 & 49.49 & 49.17 & $\uparrow98.3\%$  \\
    RAG$_{Dependency}$ & 23.00 & 27.05 & 28.57 & 30.11 & 25.16 & 28.09 & 26.99 & $\uparrow8.9\%$  \\
    Fine-tuning & 40.82 & 44.63 & 45.89 & 50.48 & 39.64 & 43.75 & 44.20 & $\uparrow78.3\%$  \\
    \midrule
    \multicolumn{9}{c}{Edit Similarity} \\
    \midrule
    Base & 58.12 & 60.78 & 62.10 & 63.64 & 57.98 & 60.60 & 60.54 & -   \\
    Random & 57.45 & 60.06 & 61.34 & 62.97 & 57.57 & 60.12 & 59.92 &   $\downarrow1.0\%$\\
    RAG$_{Sim}$-BM25 & \textbf{75.44} & \textbf{77.26} & \textbf{78.37} & \textbf{78.94} & 67.86 & 68.64 & \textbf{74.42} &  $\uparrow22.9\%$ \\
    RAG$_{Sim}$-CodeBERT & 57.48 & 60.21 & 61.63 & 62.85 & 57.29 & 60.14 &59.93  & $\downarrow1.0\%$  \\
    RAG$_{Sim}$-UniXcoder & 67.48 & 70.08 & 71.57 & 74.95 & 68.02 & 70.48 &  70.43 & $\uparrow16.3\%$\\
    RAG$_{Sim}$-CodeT5 & 68.18 & 70.49 & 71.81 & 72.87 & 68.49 & 70.56 & 70.40 &  $\uparrow16.3\%$ \\
    RAG$_{Sim}$-CoCoSoDa & 72.17 & 74.56 & 75.67 & 76.81 & \textbf{72.49} & \textbf{74.45} & 74.36 &  $\uparrow22.8\%$ \\
    RAG$_{Dependency}$ & 59.98 & 61.62 & 63.79 & 65.08 & 59.24 & 62.10 & 61.97 & $\uparrow2.4\%$  \\
    Fine-tuning & 71.52 & 74.12 & 74.82 & 77.28 & 70.10 & 72.87 & 73.45 &  $\uparrow21.3\%$ \\
    \midrule
    \multicolumn{9}{c}{BLEU Score} \\
    \midrule
    Base & 38.15 & 41.47 & 43.02 & 44.73 & 38.44 & 41.63  & 41.24 & -  \\
    Random & 37.82 & 41.03 & 42.56 & 44.37 & 38.74 & 41.20 & 40.95 & $\downarrow0.7\%$  \\
    RAG$_{Sim}$-BM25 & \textbf{66.22} & \textbf{68.35} & \textbf{69.64} & \textbf{70.34} & 58.87 & 59.99 & \textbf{65.57} & $\uparrow59.0\%$  \\
    RAG$_{Sim}$-CodeBERT & 37.93 & 41.05 & 42.79 & 44.29 & 38.33 & 41.05 & 40.91 & $\downarrow0.8\%$  \\
    RAG$_{Sim}$-UniXcoder & 54.55 & 55.77 & 57.30 & 62.22 & 52.23 & 55.19 & 56.21 & $\uparrow36.3\%$\\
    RAG$_{Sim}$-CodeT5 & 53.78 & 56.75 & 58.08 & 59.41 & 53.87 & 56.62 & 56.42 & $\uparrow36.8\%$  \\
    RAG$_{Sim}$-CoCoSoDa & 59.71 & 62.61 & 63.77 & 64.99 & \textbf{59.64} & \textbf{62.05} & 62.13 & $\uparrow50.7\%$  \\
    RAG$_{Dependency}$ & 41.27 & 44.55 & 46.26 & 47.86 & 41.10 & 44.73 & 44.29 &  $\uparrow7.4\%$ \\
    Fine-tuning & 57.02 & 60.50 & 61.69 & 65.17 & 54.82 & 58.51 & 59.62 &  $\uparrow44.6\%$ \\
    \bottomrule
    \end{tabular}
    
    \label{tab:rq1}
    \vspace{-6pt}
\end{table*}

\subsection{RQ1: Effectiveness of RAG and Fine-Tuning}
In this section, we present the performance of LCMs in our industrial code completion task in Table \ref{tab:rq1}. From the results, we can reach the following observations.

\textbf{LCMs demonstrate suboptimal performance on our industrial code completion task}. Among the six evaluated models, even the best-performing Qwen2.5-Coder 7B achieves only approximately 28\% EM and 44\% BLEU score. These relatively low metrics indicate that these models struggle with our complex, domain-specific codebase that differs substantially from their training distribution. Furthermore, we observe a clear correlation between model size and completion performance. As the parameter count increases, models demonstrate better capability in capturing code patterns, leading to better task performance. This suggests that while the domain gap poses a notable challenge, larger models' enhanced capacity allows them better to generalize their learned programming patterns to unfamiliar domains. 

\textbf{Both RAG and fine-tuning demonstrate effectiveness in enhancing model performance on industrial code completion tasks.} For \textbf{RAG} methods, our experiments show that similarity-based retrieval (RAG$_{sim}$) yields substantial improvements in completion accuracy. Specifically, when using different embedding methods for retrieval, the model's EM increases by 68.0\%, 70.8\%, and 98.3\% for UniXcoder, CodeT5, and CoCoSoDa, respectively. Surprisingly, the simple BM25 method achieves the best performance among all RAG approaches, outperforming the second-best method CoCoSoDa by 9.3\% in EM. The results suggest that in C++ code retrieval, lexical matching might be more effective than semantic matching through neural embeddings. This could be attributed to C++'s strict syntax and the prevalence of domain-specific identifiers, where exact keyword matches might be more indicative of relevance than contextual similarities captured by neural models. Moreover, we observe that RAG with CodeBERT embeddings performs poorly, showing similar results with random retrieval and even leading to slight performance degradation. This finding underscores the critical dependence of similarity-based RAG on retrieval accuracy - when the embedding model fails to effectively capture code similarities (as in the case of CodeBERT), it may adversely affect the code completion performance.

For dependency-based retrieval RAG$_{Dependency}$, we observe modest improvements of 8.9\% in EM compared to the base model. Such gains are substantially smaller than those achieved by similarity-based retrieval, suggesting that retrieving similar implementations is more beneficial than retrieving function dependencies for line-level code completion tasks.

\textbf{Fine-tuning} also demonstrates notable improvements in model performance. Specifically, after fine-tuning, the models show average improvements of 78.3\%, 21.3\%, and 44.6\% in exact match, edit similarity, and BLEU score respectively. However, we observe that the overall performance of fine-tuned models, while substantial, does not match the upper bound achieved by RAG${sim}$-BM25 and RAG${sim}$-CoCoSoDa. Nevertheless, fine-tuning does outperform RAG${sim}$ variants using other embedding models, positioning it as a competitive adaptation strategy. This comparison suggests that while fine-tuning offers consistent improvements, the quality of retrieved examples in RAG approaches can potentially lead to superior performance when optimal embedding models are employed.

\finding{1}{Our evaluation of LCMs on industrial code completion reveals that LCMs demonstrate limited effectiveness on industrial code completion, indicating a substantial domain gap. In addition, both RAG and fine-tuning substantially improve performance in all evaluation metrics. RAG approaches achieve a higher performance ceiling compared to fine-tuning, with BM25 emerging as the most effective method compared to neural embedding models.}
\vspace{-4pt}
\subsection{RQ2: Combining RAG and Fine-Tuning}
\begin{table}[t]
    \centering
    \caption{Results of combining RAG with fine-tuning, where the results are the average of six experimented LCMs.}
    \vspace{-4pt}
    \begin{tabular}{c|ccc}
    \toprule
    Methods &  EM & ES & BLEU\\
    \midrule
    Base+RAG$_{Sim}$-BM25 & 53.76& 74.42 & 65.57 \\
    FT+RAG$_{Sim}$-BM25  & \textbf{57.43} & \textbf{76.90}& \textbf{69.78} \\
    \midrule
    Base+RAG$_{Sim}$-UniXcoder & 41.64 & 70.43 & 56.21 \\
    FT+RAG$_{Sim}$-UniXcoder  & \textbf{52.08} &\textbf{77.05} & \textbf{66.45} \\
    \midrule
    Base+RAG$_{Sim}$-CodeT5 & 42.34 &70.40 & 56.42 \\
    FT+RAG$_{Sim}$-CodeT5  & \textbf{52.11} & \textbf{77.18} & \textbf{66.05} \\
    \midrule
    Base+RAG$_{Sim}$-CoCoSoDa & 49.17 & 74.36& 62.13 \\
    FT+RAG$_{Sim}$-CoCoSoDa  & \textbf{56.45} & \textbf{79.55}& \textbf{69.68} \\
    
    \bottomrule
    \end{tabular}
    \label{tab:rq1_combine}
\end{table}
To investigate whether fine-tuning can be effectively combined with RAG, we conduct experiments using different retrieval methods on fine-tuned and base models. The results are shown in Table \ref{tab:rq1_combine}. Note that the following experiments we focus on RAG$_{sim}$ due to their superior performance. 

Our experiments demonstrate that combining fine-tuning with RAG consistently yields better performance across all retrieval methods. For instance, using BM25 as the retrieval method, fine-tuned models with RAG achieve improvements of 3.67, 2.48, and 4.21 absolute percentage points in EM, ES, and BLEU scores respectively, compared to base models with RAG. This performance boost is even more pronounced with other retrieval methods. Overall, the average improvements on the three metrics reach 7.79, 5.27, and 7.91, respectively.

These consistent improvements across different retrieval methods suggest that fine-tuning and RAG complement each other effectively. The domain adaptation achieved through fine-tuning appears to enhance the model's ability to better utilize retrieved contexts, resulting in superior performance compared to using either approach independently.

\finding{2}{Combining fine-tuning and RAG creates synergistic effects in improving code completion performance. When used together, these approaches consistently outperform their individual applications, with average improvements of 7.79\%, 5.27\%, and 7.91\% in exact match, edit similarity, and BLEU scores respectively.}

\subsection{RQ3: Efficiency of RAG and Fine-Tuning}

According to previous work~\citep{DBLP:conf/sigsoft/WangHGJ0HLD23}, code completion is a time-sensitive task, leading efficiency to a crucial factor when adopting RAG and fine-tuning in this task. Therefore,
in this section, we quantitatively investigate the efficiency impact of RAG and fine-tuning approaches on code completion. Our analysis focuses on both the \textbf{preparation stage} and \textbf{runtime stage} to examine their time and resource consumption.

\subsubsection{Preparation Stage}
For the \textbf{preparation stage}, the time and resource consumption are presented in Table \ref{tab:rq2_ft_prepare}. Fine-tuning primarily consumes GPU resources, and due to the large volume of training data and model size, the training process is computationally intensive as shown in Table \ref{tab:rq2_ft_prepare}. On our 8×A100 80GB GPU server, training Qwen2.5-Coder requires 3.5 to 23.2 hours, varying by model size from 0.5B to 7B parameters, with GPU memory usage ranging from 18.3 GB to 74.8 GB. Similarly, training DeepSeek-Coder 1.3B and 6.7B consumes 6.8 and 41.4 hours respectively, utilizing 17.6 GB and 67.2 GB of GPU memory. These substantial resource requirements and A100 hours represent a remarkable bottleneck for the fine-tuning approach. 

For RAG approaches, the preparation stage primarily consists of two phases: embedding computation and index construction. As shown in Table \ref{tab:rq2_rag_prepare}, the time consumption varies greatly across different methods. BM25, being a lexical-based method, requires no embedding computation and completes index construction in merely 2 minutes. In contrast, neural embedding-based methods incur substantial computational overhead. The embedding phase, executed on a single A100 GPU, takes between 34 minutes (CodeT5) and 74 minutes (UniXcoder). The subsequent indexing phase for these neural methods requires an additional 19-37 minutes, with UniXcoder taking the longest at 37 minutes. In total, while BM25 completes preparation in 2 minutes, neural embedding-based approaches require 62-111 minutes. The variation in computation time among neural models primarily stems from their architectural differences and embedding dimensions.

\subsubsection{Runtime Stage}
For the runtime stage, the efficiency impact of RAG can be attributed to two main factors: (1) the embedding and retrieval phase of code completion queries (using embedding models), and (2) the increased query length due to the concatenation of Top-k retrieved code snippets, which subsequently affects model inference speed.

For the first factor, we measure the time consumption of both embedding and retrieval phases across our five similarity-based RAG methods, with results shown in Table \ref{tab:rq2_rag_prepare}. We find that during runtime, BM25 demonstrates superior efficiency with only 10.8ms for retrieval operations, obviously less than neural embedding-based methods. Among neural approaches, CodeBERT and UniXcoder show moderate latency (49.2ms and 46.8ms), while CodeT5 and CoCoSoDa require substantially more time (116ms and 114ms). 

\begin{figure}[t]
    \centering
    \includegraphics[width=0.95\linewidth]{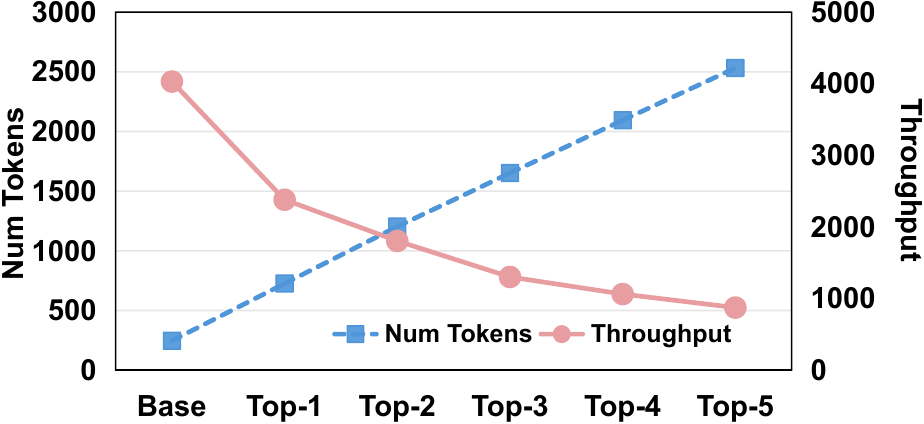}
    \vspace{-4pt}
    \caption{Average exact match and throughput under different retrieved code snippets across six LCMs.}
    \label{fig:rq2}
    \vspace{-8pt}
\end{figure}

For the second factor, we conduct an in-depth analysis using BM25 as a representative case to investigate how RAG affects query length and model inference efficiency. As shown in Figure \ref{fig:rq2}, we compare the average input length and model throughput across various scenarios: from base queries to queries augmented with top-1 through top-5 similar code snippets. The base queries have an average length of 246 tokens, but this length increases linearly with each additional retrieved code snippet, reaching approximately 2,500 tokens when incorporating top-5 similar codes. This increase in input length significantly impacts the model's throughput. Taking BM25 as an example, the average processing speed drops substantially from 4,031 tokens/second for base queries to around 874 tokens/second when including top-5 retrieved snippets.

In contrast, fine-tuning introduces no additional efficiency overhead during the runtime stage. After the resource-intensive fine-tuning process, where model parameters are adapted to our domain-specific dataset, the model can achieve substantial performance improvements using only base queries during inference. This makes fine-tuning more efficient at runtime compared to RAG, despite its higher upfront computational costs during the preparation stage.

\finding{3}{RAG and fine-tuning exhibit distinct efficiency trade-offs in code completion. Fine-tuning requires substantial computational resources (e.g., 41.4 hours and 67.2 GB GPU memory for DeepSeek-Coder 6.7B) but introduces no runtime overhead. In contrast, RAG has minimal preparation costs but incurs runtime overhead through context retrieval and increased input lengths, reducing model throughput by up to 78\% (from 4,031 to 874 tokens/second) when using top-5 retrieved snippets.}


\begin{table}[t]
    \centering
    \caption{Preparation (training) time and resource consumption of fine-tuning different LCMs.}
    \vspace{-6pt}
    \resizebox{\linewidth}{!}{
    \begin{tabular}{c|cccc|cc}
    \toprule
    & QC-0.5B & QC-1.5B & QC-3B &QC-7B &DSC-1.3B & DSC-6.7B  \\
     \midrule
   Time & 3.5h & 4.6h & 10.1h & 23.2h & 6.8h & 41.4h \\
   Memory & 18.3G & 31.5G & 44.5G & 74.8G & 17.6G & 67.2G\\
    \bottomrule
    \end{tabular}
    }
    \vspace{-6pt}
    \label{tab:rq2_ft_prepare}
\end{table}

\begin{table}[t]
    \centering
    \caption{Preparation (embedding and indexing) and retrieval time consumption of RAG methods.}
    \vspace{-6pt}
    \resizebox{\linewidth}{!}{
    \begin{tabular}{c|c|cccc}
    \toprule
      Stage & BM25 & CodeBERT & UniXcoder & CodeT5 & CoCoSoDa  \\
      \midrule
     Embedding  & - & 53min & 74min & 34min & 64min \\
     Indexing & 2min  & 19min  & 37min & 28min & 35min \\
     \midrule
     Runtime Retrieval& 10.8ms  & 49.2ms & 46.8ms &116ms &  114ms \\ 
     \bottomrule
    \end{tabular}
    }
    \vspace{-6pt}
    \label{tab:rq2_rag_prepare}
\end{table}

\subsection{RQ4: Parameter Analysis}

\begin{figure*}[t]
    \centering
    \includegraphics[width=0.95\linewidth]{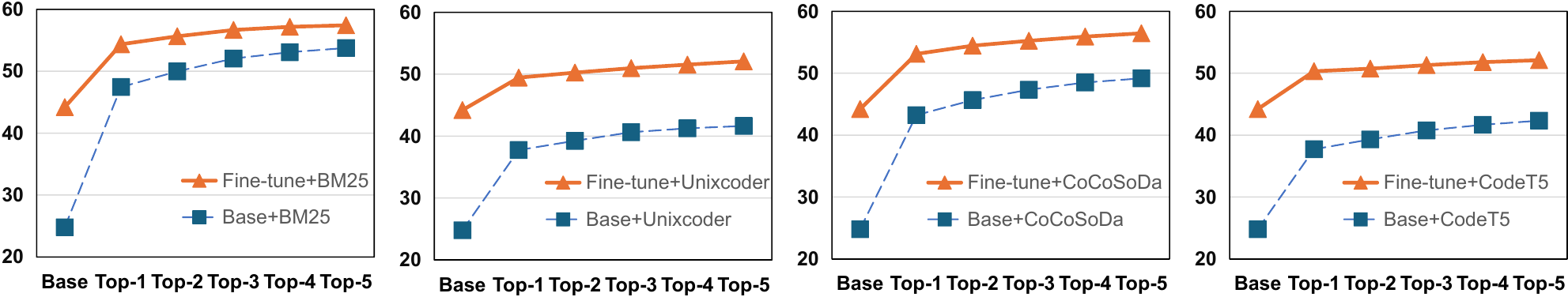}
    \vspace{-6pt}
    \caption{Exact match of the base and fine-tuned model under different retrieved contexts.}
    \label{fig:rq4_2}
    \vspace{-6pt}
\end{figure*}
\subsubsection{Impact of Retrieved Context Size}
We conduct a detailed analysis of how base models and fine-tuned models perform when combined with RAG, specifically examining performance changes as the number of retrieved contexts (top-k) increases. We exclude CodeBERT from this analysis due to its poor performance in selecting relevant contexts. As shown in Figure \ref{fig:rq4_2}, both base models and fine-tuned models demonstrate performance improvements as the number of retrieved contexts (K) increases. However, the improvement patterns differ significantly between them. While fine-tuned models combined with RAG achieve higher absolute performance, the marginal benefit (slope of improvement) from additional retrieved contexts is less pronounced compared to base models with RAG.

\begin{figure*}[t]
    \centering
    \includegraphics[width=0.8\linewidth]{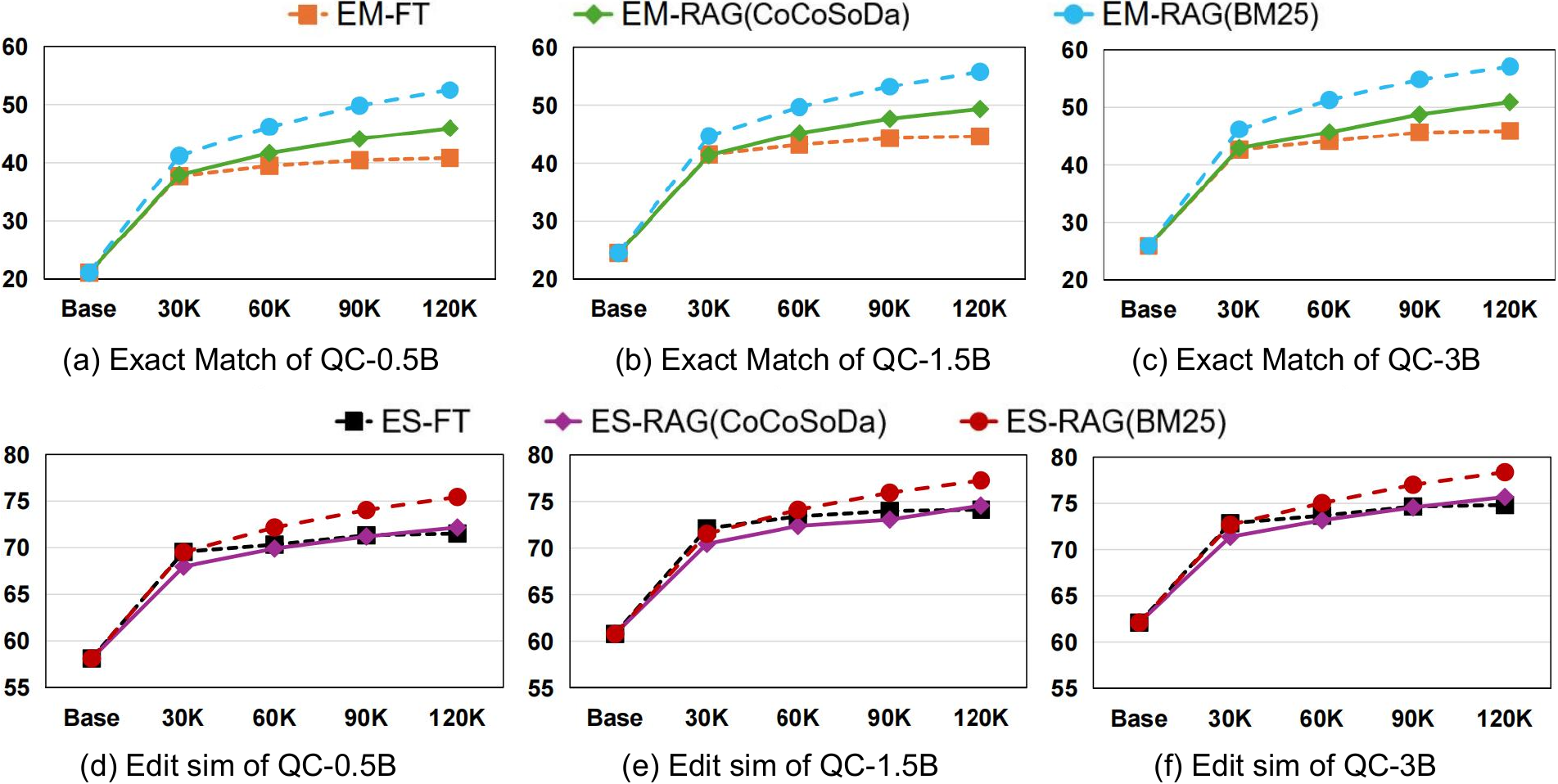}
    \vspace{-4pt}
    \caption{Exact match and edit similarity of fine-tuning and RAG under different sizes of codebase.}
    \label{fig:rq4}
    \vspace{-12pt}
\end{figure*}

\subsubsection{Impact of the Scale of CodeBase}
Both RAG and fine-tuning require a codebase during their adaptation process, i.e., either as a retrieval database or training dataset. In this research question, we investigate how sensitive these approaches are to the scale of the codebase. We randomly sample different sizes of C++ files (30k, 60k, 90k, and 120k) from our codebase to create varying-sized retrieval databases and training sets. Due to the computational intensity of training and evaluation, we conduct experiments using Qwen models of smaller sizes (0.5B, 1.5B, and 3B parameters). For RAG, we focus on BM25 and CoCoSoDa as retrieval methods, as they demonstrate superior performance in our previous experiments. The results of these experiments are presented in Figure \ref{fig:rq4}.

The experimental results reveal that both fine-tuning and RAG exhibit performance improvements as the codebase size increases, with the most substantial gains observed when transitioning from the base model to the 30K files included. Specifically, across the three model sizes, fine-tuning and RAG-BM25 demonstrate average relative improvements of 64.6\% and 78.3\% in EM, respectively. As the codebase expands further, the performance gains gradually diminish, but with notably different patterns between fine-tuning and RAG. When scaling from 90K to 120K files, fine-tuning shows diminishing returns, with average improvements of merely 0.35\% and 0.16\% percentage points in exact match and edit similarity across the three models. In contrast, RAG methods maintain more substantial improvements: RAG-BM25 and RAG-CoCoSoDa achieve average EM gains of 2.26\% and 2.13\% across all model sizes. This phenomenon suggests that fine-tuning approaches a performance plateau in our dataset when the training data exceeds approximately 300M tokens (90K files). However, RAG methods continue to derive meaningful benefits from larger-scale codebases, demonstrating better scalability in leveraging additional code resources.

\finding{4}{Both RAG and fine-tuning demonstrate sensitivity to codebase scale but with distinct scaling patterns. While both approaches show substantial initial gains, fine-tuning exhibits diminishing returns beyond 90K files with minimal improvements. In contrast, RAG methods maintain meaningful performance gains even at larger scales (90K to 120K files), suggesting superior scalability in leveraging expanded codebases for code completion.}
\section{Discussion}
\subsection{Impact of Fine-tuning on Model's Generalization Ability}

In this section, we evaluate the catastrophic forgetting effect on generalization ability induced by fine-tuning, specifically examining how the model's performance on other tasks and benchmarks is affected after fine-tuning on our code repository. We select three representative tasks: code generation, code reasoning, and code translation. For code generation, following the work~\citep{DBLP:conf/icml/0003W0D024,DBLP:conf/iclr/LuoX0SGHT0LJ24}, we evaluate the HumanEval and MBPP datasets using pass@1 as the evaluation metric. For code reasoning, we employ the CRUX-Eval dataset~\citep{DBLP:conf/icml/GuRLSS024}, where the model is required to infer either the output given code and input (CRUX-O), or the input given code and output (CRUX-I), also measured by pass@1. For code translation, we utilize the CodeXGLUE benchmark \citep{lu1codexglue}, measuring performance with the CodeBLEU metric \citep{ren2020codebleu}. The experimental results are presented in Table \ref{tab:rq3}, which demonstrates the model's performance on these general tasks before and after fine-tuning on our domain-specific dataset.

\begin{table*}[t]
    \centering
    \caption{Comparison of base and fine-tuned models on the code generation, code reasoning, and code translation tasks.}
    \vspace{-6pt}
    \begin{tabular}{c|cc|cc|c|c|c|c}
    \toprule
    \multirow{2}{*}{Models} & HumanEval & +Plus & MBPP & +Plus & CRUX-O & CRUX-I & C\#2Java & Java2C\# \\
         & \multicolumn{6}{c|}{Pass@1} & \multicolumn{2}{c}{CodeBLEU} \\
    \midrule
    QC-0.5B & 26.83 & 22.56 & 50.88 & 44.11 & 23.62 & 22.25 & 41.71& 32.79 \\
    Fine-tuned & 18.29 & 15.24 & 36.09 & 29.82 & 21.25& 20.00 &40.57 & 33.83 \\
    \hline
    QC-1.5B & 46.34 & 37.80 & 65.16 & 53.13 & 35.00 & 25.12 & 45.95& 36.15 \\
    Fine-tuned & 21.95 & 17.68 & 61.65 & 49.87 & 30.50& 23.12 & 44.90&34.94  \\
    \hline
    QC-3B & 49.39 & 39.02 & 67.17 & 53.88 & 39.12& 30.75 & 47.61& 37.19 \\
    Fine-tuned & 40.85 & 32.32 & 64.66 & 51.88 &36.50 & 28.38 &47.32 & 36.70 \\
    \hline
    QC-7B & 62.20 & 52.44 & 70.43 & 57.89 & 50.12&  36.00& 48.00&38.92  \\
    Fine-tuned & 45.73 & 40.24 & 36.09 & 29.82 & 45.63& 33.38 & 47.52& 38.11 \\
    \hline
    DSC-1.3B & 31.10 & 25.61 & 55.89 & 45.86 & 29.88 & 24.75 & 44.80 &31.22  \\
    Fine-tuned & 29.27 & 23.78 & 55.39 & 44.36 & 28.25& 22.62 &43.93 &34.57 \\
    \hline
    DSC-6.7B & 52.44 & 45.12 & 64.91 & 51.63 & 42.12& 33.38 & 46.24&34.29  \\
    Fine-tuned & 47.56 & 39.63 & 62.16 & 49.62 & 43.38& 35.38 &45.88 &35.33  \\
    \midrule
    Avg. Base & 44.72 & 37.04 & 62.41 & 51.08 & 36.64 & 28.71 & 45.72 & 35.09 \\
    Avg. Fine-tuned & 33.95 & 28.15 & 55.14  & 42.56& 34.25& 27.15 & 45.02 & 35.60\\
    \bottomrule
    \end{tabular}
    \vspace{-8pt}
    \label{tab:rq3}
\end{table*}

Based on our experimental results, we observe varying degrees of performance degradation across different tasks, indicating a nuanced pattern of catastrophic forgetting~\citep{DBLP:conf/icse/GaoZGW23,mccloskey1989catastrophic}. Specifically, in code generation tasks, we observe average Pass@1 decreases of 24.1\% and 11.7\% on HumanEval and MBPP respectively. When evaluated with stricter test cases (+Plus), this performance gap widens to 24.0\% and 16.7\%.

Regarding code reasoning and translation capabilities, we observe a much milder impact from fine-tuning compared to code generation tasks. For code reasoning, fine-tuned models show average decreases of 6.5\% and 5.8\% on CRUX-O and CRUX-I respectively. In code translation tasks, the impact is even smaller, with only a marginal decrease of 0.7\% in C\#2Java and, notably, a slight improvement of 0.5\% in Java2C\# after fine-tuning.

This pattern suggests that fine-tuning our dataset has a minimal negative impact on the model's reasoning and translation capabilities, starkly contrasting to the more substantial degradation observed in code generation tasks. Particularly encouraging is the case of Java2C\# translation, where fine-tuned models actually outperform their base counterparts on average. We also observe a positive outlier with DeepSeek-Coder 6.7B, which shows improvements in both reasoning (3.0\% and 6.0\% on CRUX-O and CRUX-I respectively) and translation tasks after fine-tuning. These findings suggest that fine-tuning has varying impacts across different code-related capabilities. While code generation performance shows notable degradation, the effects on code reasoning and translation abilities are much less pronounced. 

\subsection{Implication of Findings}
In this section, we discuss the implications of our work for developers and researchers.

\subsubsection{For Developers}
Our main results demonstrate that RAG and fine-tuning present distinct trade-offs in effectiveness and efficiency for practical applications. Based on these findings, we offer the following implications for developers adapting LLMs in industrial settings:

(1) \textbf{Exploring fine-grained source code segmentation for RAG}: Our experiments, which used function-level retrieval, showed that adding a single relevant code snippet during inference increases input tokens by approximately 500 on average, which remarkably affects inference speed. Investigating finer-grained RAG approaches that reduce token length while maintaining retrieval effectiveness could lead to more efficient solutions. This might include exploring approaches such as snippet-level or statement-level retrieval.

(2) \textbf{Selecting RAG or fine-tuning based on available resources}: RAG and fine-tuning consume computational resources at different stages of deployment. While RAG requires more computational resources during runtime, fine-tuning demands substantial resources during the preparation stage. Given that training and inference typically utilize different GPU configurations \citep{jiang2018efficient, reuther2022ai}, developers should select their adaptation approach based on their available resource distribution. In resource-rich environments, combining both approaches can yield optimal results.

(3) \textbf{Using RAG for ensuring LCMs' generalization ability}: 
When deploying models that need to handle diverse scenarios, RAG emerges as the preferred choice. Our experiments show that fine-tuning can compromise a model's generalization abilities such as in code generation and reasoning tasks. Therefore, when the deployment scenario requires maintaining both domain-specific performance and general-purpose capabilities, RAG presents a more balanced solution.

(4) \textbf{Selecting RAG in data-rich scenarios}: Our parametric studies reveal that fine-tuning's performance improvements plateau as the codebase scale increases, while RAG continues to show significant performance gains. In scenarios with abundant code data, RAG demonstrates superior scalability and can provide larger performance gains. This suggests that organizations with large proprietary codebases might find RAG particularly advantageous for their code completion systems.

\subsubsection{For Researchers}
In Section \ref{sec:results}, we demonstrate that both fine-tuning and RAG are effective approaches for adapting LLMs to proprietary codebases, substantially improving code completion performance on private datasets. However, our findings reveal that each approach has distinct advantages and limitations, suggesting several promising research directions:

(1) \textbf{Exploring more advanced code embedding models}: Our experiments show that RAG's effectiveness is directly correlated with retrieval performance. Investigating more sophisticated code embedding models could lead to more relevant code retrieval and, consequently, better completion accuracy. This includes developing embedding techniques that better capture semantic relationships and domain-specific code patterns.

(2) \textbf{Exploring forgetting-aware fine-tuning techniques}: While fine-tuning effectively improves domain-specific performance, we observe degradation in general capabilities, particularly in code generation tasks. Research into more robust fine-tuning techniques that maintain the model's general capabilities while enhancing domain-specific performance represents a crucial direction. This could involve exploring methods such as elastic weight consolidation~\citep{kirkpatrick2017overcoming} or knowledge distillation~\citep{DBLP:journals/corr/HintonVD15}.

(3) \textbf{Optimizing long-context inference}: We observe that inference speed is significantly affected by token length, particularly in industrial codebases which typically involve long and complex source code. Research into optimizing long-sequence handling, such as efficient attention mechanisms or context compression techniques, is crucial for improving the practical efficiency of code completion systems in industrial settings.

\subsection{Case Study}
\begin{figure}[t]
    \centering
    \includegraphics[width=1\linewidth]{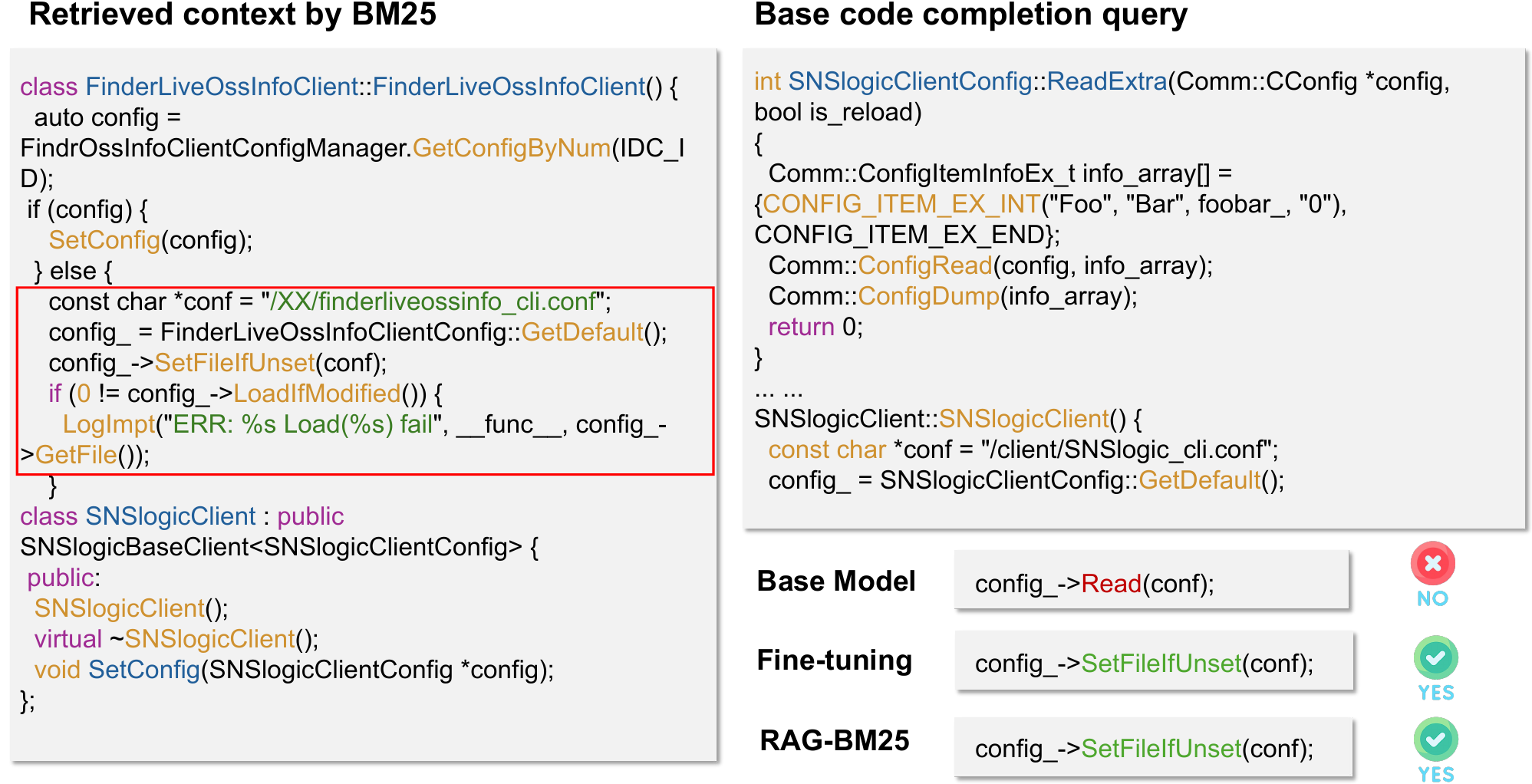}
    \vspace{-8pt}
    \caption{Example that the QC-7B base model fails to predict the next line of code but fine-tuning and RAG succeed.}
    \label{fig:case1}
    \vspace{-6pt}
\end{figure}
In this section, we conduct case studies to illustrate how fine-tuning and RAG influence LLM's code completion predictions in practice.

Figure \ref{fig:case1} demonstrates the varying completion results across different approaches. For the base code completion query (shown on the right), the base QC-7B model fails to predict the correct config-related operation, generating a plausible but incorrect \texttt{Read(conf)} call. This limitation stems from the model's lack of knowledge about the proprietary codebase's specific APIs and conventions.

After fine-tuning, the model successfully generates the correct code snippet, specifically using the \texttt{SetFileIfUnset} method. This improvement can be attributed to the model's acquired knowledge about \texttt{SNSLogicClientConfig} and its associated usage patterns through training on the proprietary codebase.

Similarly, the RAG-BM25 approach also generates the correct completion. The success of RAG can be traced to its retrieval mechanism, which identifies a relevant code snippet from another \texttt{Client} class (highlighted in red in the retrieved context). This retrieved code example contains similar config operations, providing a valuable reference for the current prediction task. This demonstrates how RAG effectively leverages existing codebase knowledge to guide accurate code completion.
\subsection{Threat to Validity}
\textbf{External Validity}: Our study primarily relies on an industrial codebase from our collaborating company, which may exhibit company-specific patterns and practices that differ from other software organizations. To mitigate this limitation, we deliberately collect a diverse set of code samples across different functional domains and business units within the company. Our dataset encompasses over 160,000 code files, providing a comprehensive representation of the department's development practices and coding patterns.

\textbf{Internal Validity}: The efficiency measurements in our study are conducted under our available computational configurations. The relative performance characteristics of RAG and fine-tuning approaches might vary under different resource constraints. Specifically, the efficiency trade-offs could shift with varying training resources (GPU configurations), inference resources (deployment environments), or CPU resources (for retrieval operations). This suggests that our efficiency results should be interpreted within the context of our experimental setup.
\section{Related Work}
\subsection{Retrieval-Augmented Generation}
Retrieval-augmented generation (RAG) is an approach that improves generation quality by incorporating information from external knowledge sources. This approach enhances the accuracy and reliability of LLMs' outputs and is widely used in code-related tasks~\citep{DBLP:conf/acl/LuDHGHS22,DBLP:conf/sigsoft/Wang0JH23}. For example, Reacc~\citep{DBLP:conf/acl/LuDHGHS22} leverages both lexical copying and referring to code with similar semantics to improve code completion. Parvez et al.~\citep{DBLP:conf/emnlp/ParvezACRC21} propose REDCODER which enhances code generation and summarization by retrieving and leveraging relevant code snippets and summaries from a reference database. Similarly, Wang et al.~\citep{DBLP:conf/sigsoft/Wang0JH23} and Peng et al.~\citep{DBLP:conf/icse/PengGGHL24} advanced automatic program repair by explicitly utilizing fix templates from historical bug-fix pairs.

\subsection{Large Code Models}
Recently, Large Code Models have emerged as powerful tools for diverse software engineering tasks~\citep{DBLP:conf/emnlp/ZhangCZKLZMLC23,DBLP:conf/icml/0003W0D024,DBLP:conf/kbse/WangLPGCWGL23,ma2023oops,li2023protecting,ji2023benchmarking,wong2023refining,wang2024exploring,li2024api,ji2024testing,zhang2025low}. Various influential models have demonstrated significant capabilities in code generation and understanding. For example, Meta's Code Llama~\citep{roziere2023code} is a foundation model that builds upon the original LLaMA architecture and extends the context window to 16K tokens and serves as a versatile foundation model for programming tasks. DeepSeek-Coder\citep{guo2024deepseek} is a series of large code models that have an architecture identical to CodeLlama. DeepSeek-Coder is trained from 2T tokens from scratch, achieving state-of-the-art performance in a variety of code intelligence tasks. Qwen2.5 Coder~\citep{hui2024qwen2} is a family of LCMs based on models of the Qwen2.5 series. Qwen2.5 undergoes four-stage training with a total number of 5.5T tokens, presenting state-of-the-art performance in both code generation and reasoning.

\subsection{Code Completion}
Code completion predicts subsequent code tokens or statements to assist programmers during software development~\citep{DBLP:conf/icse/LiWLWCWG23,li2022cctest,li2022unleashing,li2023feasibility}. With the development of deep learning, various DL-based code completion methods are applied to code completion and achieve state-of-the-art performance. For instance, Li et al.~\citep{DBLP:conf/ijcai/LiWLK18} proposed a point-mixture network to relieve the out-of-vocabulary problem in sequence-to-sequence code completion. Izadi et al.~\citep{DBLP:conf/icse/IzadiGG22} proposed CodeFill, which integrates both type and semantic information to improve the precision of completion. Recent research has increasingly focused on leveraging repository-level context. For example, Repocoder~\citep{DBLP:conf/emnlp/ZhangCZKLZMLC23} utilizes useful information scattered in different files by iterative completion and retrieval. Liu et al.~\citep{DBLP:conf/iclr/0003XM24} proposed RepoBench to evaluate code completion systems under complex and multi-file programming scenarios. RLcoder~\citep{DBLP:journals/corr/abs-2407-19487} employs RAG and trains the retriever to learn to retrieve useful content from the repository. 
\section{Conclusion}
In this paper, we conduct extensive experiments to evaluate the effectiveness of RAG and fine-tuning in LCM-based code completion in the industry. We reveal the effectiveness and efficiency trade-offs when adopting the two methods: RAG achieves a higher performance ceiling while introducing non-trivial runtime completion overhead. Our findings provide guidance for both researchers and developers for further research directions and LCM adoption.

\section*{Acknowledgment}
The work described in this paper was supported by the Research Grants Council of the Hong Kong Special Administrative Region, China (No. SRFS2425-4S03 of the Senior Research Fellow Scheme). This work is also supported by the CCF-Huawei Populus Grove Fund. 

\bibliographystyle{ACM-Reference-Format}
\bibliography{sample-base}

\end{document}